\documentclass[referee]{aa}

\usepackage{natbib}
\bibpunct{(}{)}{;}{a}{}{,} 

\usepackage{txfonts}

\usepackage{graphicx}

\newcommand{\OO}[1]{{\cal O}(c^{-#1})}
\newcommand{\vecg}[1]{\mbox{\boldmath$#1$}}
\newcommand{\ve}[1]{\vecg{#1}}
\newcommand{\muas}[0]{\hbox{\rm $\mu$as}}

\begin{document}

\title{Numerical Simulations of the Light Propagation in
the Gravitational Field of Moving Bodies}

\author{Sergei A. Klioner \and Michael Peip}

\institute{Lohrmann Observatory, Dresden Technical University,
Mommsenstr. 13, 01062 Dresden, Germany}

\offprints{Sergei A. Klioner, \email{klioner@rcs.urz.tu-dresden.de}}

\date{Received \today  / Accepted \today}

\abstract{
One of the most subtle points in the modern relativistic models for
microarcsecond astrometrical observations is the treatment of the
influence of translational motion of gravitating bodies on the light
propagation. This paper describes numerical simulations of the light
propagation in the gravitational field of moving gravitating bodies
as well as summarizes the underlying theory. The simulations include
high-precision numerical integrations of both post-Newtonian and
post-Minkowskian differential equations of light propagation and a
detailed comparison of the results of the numerical integrations with
various available approximate analytical formulas. The simulations has
been performed both for hypothetical bodies with various parameters of
trajectories as well as for all the major bodies of the solar system
using the JPL ephemeris DE405/LE405 to calculate their motion.

It is shown that for the accuracy of $\sim 0.2$~\muas\ it is sufficient
to use the well-known solution for the light propagation in the field
of a motionless mass monopole and substitute in that solution the
position of the body at the moment of closest approach between the
actual trajectory of the body and the unperturbed light path (as it was
first suggested by \citet{Hellings:1986}). For a higher accuracy one
should use either the post-Newtonian solution for uniformly moving
bodies \citep{Klioner:Kopeikin:1992} or the post-Minkowskian solution
for arbitrarily moving bodies \citep{Kopeikin:Schaefer:1999}. For
astrometric observations performed from within the solar system these
two solutions guarantee the accuracy of $\sim~0.002$~\muas\ and
are virtually indistinguishable from each other.
}

\titlerunning{Numerical Simulations of the Light Propagation}
\authorrunning{S.A.~Klioner, M.~Peip}

\maketitle

\section{Introduction}

It is widely known that extremely high accuracy of the future
space-born astrometric missions like GAIA
\citep{GAIA:2000,Perryman:et:al:2001,Bienayme:Turon:2002} and SIM
\citep{SIM:1998} makes it necessary to formulate the reduction model of
positional observations in a form fully consistent with General
Relativity Theory (GRT). The relativistic models of positional
observations has been formulated by several groups of authors:
\citet{Klioner:1989}, \citet{Brumberg:Klioner:Kopejkin:1990},
\citet{Klioner:Kopeikin:1992},
\citet{deFelice:Lattanzi:Vecchiato:Bernacca:1998},
\citet{deFelice:Vecchiato:Bucciarelli:Lattanzi:Crosta:2000},
\citet{deFelice:Bucciarelli:Lattanzi:Vecchiato:2001},
\citet{Klioner:2003a}. This paper is devoted to an investigation of one
subtle point in any microarcsecond relativistic model of positional
observations. Namely, the influence of the translational motion of
gravitating bodies on the light propagation is investigated here in
great detail.

After the pioneering work of \citet{Hellings:1986} where it was
suggested to compute the positions of the gravitating bodies at the
moment of closest approach between the body and the unperturbed light
ray and substitute these positions into the well-known solution for the
light propagation in the gravitational field of a system of motionless
bodies, several authors have succeeded to formulate more rigorous
approaches to the problem
\citep{Klioner:1989,Klioner:1991,Klioner:Kopeikin:1992,Kopeikin:Schaefer:1999,
Kopeikin:Mashhoon:2002}.
Detailed historical overviews can be found in Introduction of
\citet{Kopeikin:Schaefer:1999} and in Section 6 of \citet{Klioner:2003a}
(see also \citet{Klioner:2003b}).

In this paper we perform extensive numerical simulations aimed at
clarifying the ability of various approximate analytical formulas to
reproduce the gravitational light deflection in the field of the solar
system at the level of $0.1-1$ \muas\ as required by GAIA and SIM.

Possible ways to compute the light trajectory in the gravitational
field of moving bodies are summarized in
Section \ref{Section-ways}. Section \ref{Section-simulations}
 explains general layout of the
simulations. The results of the simulations are discussed in Section
\ref{Section-results}. The suggestions for practical relativistic
modeling of high-accuracy positional observations depending on the goal
accuracy are formulated in Section \ref{Section-conclusion}. In the
Appendices we summarize the theoretical formulas concerning the
influence of the translational motion of the gravitating bodies on
light propagation. Most of these formulas are used in the simulations.
Appendix \ref{Section-null-geodesics} contains general formulas
for null geodesics in the weak-field approximation. Both
post-Newtonian and post-Minkowskian equations are given there. The most
important theoretical results for light propagation in the
post-Newtonian approximation are given in Appendix \ref{Section-pN}.
The equations of light propagation in the post-Minkowskian
approximation are discussed in Appendix \ref{Section-pM}.
The two point boundary problem for analytical solutions is
discussed in Appendix \ref{Appendix-boundary-problem}.


\section{Possible ways to calculate the light propagation in the field
of moving bodies}
\label{Section-ways}

According to the theory described in the Appendices there are several
ways to calculate the light trajectory in the gravitational field of
moving mass monopoles:

\begin{enumerate}
\item Numerical integration of the post-Minkowskian differential
equations of light
propagation (\ref{eqm-photon-pM:explicit})--(\ref{eta})
with initial conditions (\ref{photon:initials-pM})--(\ref{theta}).

\item Analytical post-Minkowskian solution
(\ref{photon:solution:pM})--(\ref{dot-x-pM}) for arbitrarily moving
bodies (the solution for the position of the photon
(\ref{photon:solution:pM}) contains an integral that can be computed
numerically, or estimated to be negligible for a particular purpose and
thus neglected).

\item Numerical integration of the post-Newtonian equations of light
propagation (\ref{BRS:eqm:photon:motion:new})--(\ref{delta-pN})
with initial conditions (\ref{BRS:eqm:photon:initials})--(\ref{s-pN}).

\item Analytical post-Newtonian solution for uniformly moving mass
monopoles (\ref{BRS:photon:solution:rectilinear})--(\ref{J-dot}) with
two free parameters $\ve{x}_{A0}$ and $\ve{v}_A$ to be related to the
actual trajectory of the body.
\end{enumerate}

As discussed in the Appendices below the post-Minkowskian approximation
scheme deals with expansions in powers of the gravitational constant
$G$.
Velocities of gravitating bodies are not considered as small in the
post-Minkowskian approximation scheme which is sometimes called
weak-field fast-motion approximation.
The first post-Minkowskian approximation implies that all terms of
order ${\cal O}(G^2)$ are neglected. The post-Newtonian approximation
scheme operates with expansions in powers of $c^{-1}$.
In the post-Newtonian approximation scheme velocities of gravitating
bodies are considered to be small. This approximation scheme is
sometimes called weak-field slow-motion approximation.
In the first post-Newtonian approximation terms of order ${\cal
O}(c^{-4})$ are neglected in the equations of light propagation. One
can prove that in the case of light propagation the formulas of the
first post-Newtonian approximation are linear in $G$ and, therefore,
contained in those of the first post-Minkowskian approximation
(see e.g. Appendix \ref{Section-pM}).

The most accurate way to calculate the light propagation
is clearly the first one, i.e. numerical
integration of the post-Minkowskian differential equations of motion
for the photon. The errors of those differential equations of motion
come from the effects of the second post-Minkowskian (or
post-Newtonian) approximation which are known to be negligible for the
solar system applications. Numerical integration of the post-Newtonian
equations of motion for the photon
(\ref{BRS:eqm:photon:motion:new})--(\ref{delta-pN}) can also be used to
calculate the light trajectory in the field of arbitrarily moving
bodies. However, since the post-Newtonian equations of motion are
contained in the post-Minkowskian equations of motion (Appendix
\ref{Section-pM}), the former can be used for an internal consistency
check of the whole calculation rather than for an independent
computation of the light path. Our numerical experiments show that for
solar system applications the results of numerical integrations
of the post-Newtonian and post-Minkowskian equations of motion are
identical within the errors of the first post-Minkowskian
approximation.

The analytical post-Minkowskian solution given in Appendix
\ref{Section-pM-analytical} is surely the most accurate analytical
solution for the problem. However, (1) the solution for the photon's
position involves an integral which should be in principle computed
numerically, and (2) the post-Minkowskian solution is relatively
expensive as far as the computing time is concerned since it contains
the retarded moment of time to be computed by iterations (see below).
On the other hand, the full accuracy of the analytical post-Minkowskian
solution is not necessary to attain the accuracy of 1 \muas\ for the
solar system applications. Simpler analytical solutions of
sufficient accuracy can be found instead.

The fully analytical post-Newtonian solution given in Appendix
\ref{Section-pN-analytical-moving} describes the light trajectory
in the field of uniformly moving gravitating bodies having the
coordinates

\begin{equation}
\label{position-pN}
\ve{x}_A(t)=\ve{x}_{A0}+\ve{v}_A\,(t-t_0),
\end{equation}

\noindent
where $\ve{x}_{A0}$ and $\ve{v}_A$ are two arbitrary constant vectors. These
constants can be related to the actual trajectory of the body
$\ve{x}_A^{\rm eph}(t)$ in different ways with the hope that the errors
related to the non-uniformity of the body's trajectory will be
minimized in some sense. The principal goal of this paper is to check
if the analytical post-Newtonian solution with some reasonable choice
of the constants $\ve{x}_{A0}$ and $\ve{v}_A$
can describe the gravitational light deflection
with an accuracy of $0.1-1$ \muas.

\begin{figure*}
\sidecaption
\includegraphics[width=12cm]{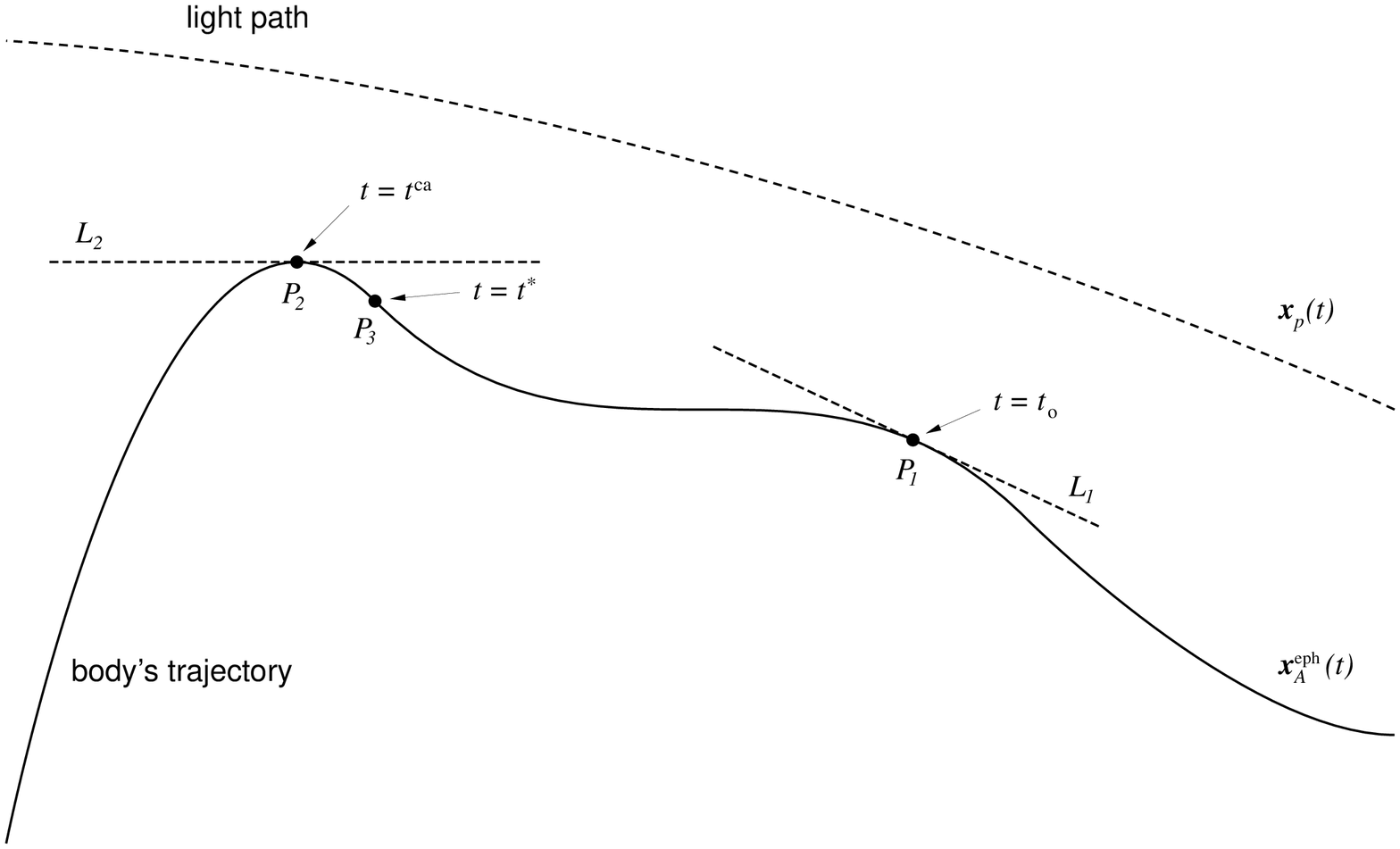}
\caption{Actual trajectory $\ve{x}^{\rm eph}_A(t)$
of body $A$ and the five simplified trajectories used to model the
light propagation within the post-Newtonian approximation scheme: three
fixed positions $P_1$ ($\ve{x}_A(t)=\ve{x}^{\rm eph}_A(t_o)={\rm
const}$), $P_2$ ($\ve{x}_A(t)=\ve{x}^{\rm eph}_A(t^{\rm ca})={\rm
const}$) and $P_3$ ($\ve{x}_A(t)=\ve{x}^{\rm eph}_A(t^*)={\rm
const}$), and two trajectories with constant velocity $L_1$
($\ve{x}_A(t)=\ve{x}^{\rm eph}_A(t_o)+\dot{\ve{x}}^{\rm
eph}_A(t_o))\,(t-t_o)$ and $L_2$ ($\ve{x}_A(t)=\ve{x}^{\rm
eph}_A(t^{\rm ca})+ \dot{\ve{x}}^{\rm eph}_A(t^{\rm ca}))\,(t-t^{\rm
ca})$. The position $P_3^\prime$ mentioned in the text is not shown on
the sketch. It is situated on the actual trajectory of the body close
to $P_3$.}
\label{figure-5-traj}
\end{figure*}

For the analytical post-Newtonian solution one has to choose either a
fixed point or a straight line as the model trajectory $\ve{x}_A(t)$ of
the body which should be distinguished from
the actual trajectory of the body $\ve{x}_A^{\rm eph}(t)$.
In this paper we consider six choices for the constants
$\ve{x}_{A0}$ and $\ve{v}_A$ in the post-Newtonian solution: four
solutions for a body at rest named $P_1$, $P_2$, $P_3$ and $P_3^\prime$
and two solutions for a body moving with constant velocity $L_1$ and
$L_2$ (see Figure \ref{figure-5-traj}). Each of the considered
solutions uses the actual position (and possibly velocity) of the body at
some reference moment of time $t=t_{\rm ref}$. The most simple solution
$P_1$ uses the fixed position $\ve{x}_A^{\rm eph}(t_o)$ of the body at
the moment of observation $t_{\rm ref}=t_o$. The fixed position
$\ve{x}_A^{\rm eph}(t^{\rm ca})$ of the body at the moment $t_{\rm
ref}=t^{\rm ca}$ of closest approach between the body and the
unperturbed light ray is used for the second solution $P_2$. The
moment $t^{\rm ca}$ can be calculated as

\begin{eqnarray}\label{moment-of-closest-approach}
t^{\rm ca}&=&\max\left(t_e,t_o-
\max\left(0,{\ve{g}\cdot (\ve{x}_p(t_o)-\ve{x}^{\rm eph}_A(t_o))
             \over c\,|\ve{g}|^2}\right)
\right),
\\
\label{vector-g}
\ve{g}&=&\vecg{\mu}-{1\over c}\,\dot{\ve{x}}^{\rm eph}_{A}(t_o),
\end{eqnarray}

\noindent
where $t_e$ is the moment of emission of the photon, and $\ve{x}_p(t)$
is the (unperturbed) trajectory of the photon. If the source is
situated outside of the solar system one can put $t_e=-\infty$ and the
outer $\max$ can be omitted in (\ref{moment-of-closest-approach}). The
retarded moment of time $t^*$ is used as $t_{\rm ref}$ in the solution
$P_3$. This moment of time is defined by

\begin{equation}\label{t-r}
t^*+{1\over c}\,|\ve{x}_p(t)-\ve{x}^{\rm eph}_A(t^*)|=t.
\end{equation}

\noindent
The moment $t^*$ is relatively expensive to calculate since the
equation (\ref{t-r}) is an implicit one and one has to use some kind of
iterations to solve it (e.g., Newton's method). That is why, one can
try to substitute $t^*$ by its simplified version $t^{*\prime}$
which can be directly calculated

\begin{equation}\label{t-r-s}
t^{*\prime}=t-{1\over c}\,|\ve{x}_p(t)-\ve{x}^{\rm eph}_A(t)|.
\end{equation}

\noindent
The solution $P_3^\prime$ uses the position $\ve{x}_A^{\rm
eph}(t^{*\prime})$ of the body at $t_{\rm ref}=t^{*\prime}$.

The solutions $L_1$ and $L_2$ are the solutions for a body moving with
a constant velocity. In these two cases the parameters $\ve{x}_{A0}$
and $\ve{v}_A$ of
(\ref{BRS:photon:solution:rectilinear})--(\ref{J-dot}) are chosen to
coincide with the actual position and velocity of the body at the
moments $t_{\rm ref}=t_o$ and $t_{\rm ref}=t^{\rm ca}$, respectively
(therefore, the trajectories of the bodies used in $L_1$ and $L_2$ are
tangents to the actual trajectory of the body at these two moments of
time). The choice of the trajectories $\ve{x}_A(t)$ are summarized in
Table \ref{Table-pN-solutions} and Figure \ref{figure-5-traj}.

\begin{table}
\tabcolsep=10pt
\begin{tabular}{llll}
solution & $t_{\rm ref}$ & $\ve{x}_{A0}$ & $\ve{v}_A$ \\
\hline
$P_1$      & $t_o$         & $\ve{x}_{A}^{\rm eph}(t_o)$ & 0 \\
$P_2$      & $t^{\rm ca}$  & $\ve{x}_{A}^{\rm eph}(t^{\rm ca})$ & 0 \\
$P_3$      & $t^*$  & $\ve{x}_{A}^{\rm eph}(t^*)$ & 0 \\
$P_3^\prime$      & $t^{*\prime}$  & $\ve{x}_{A}^{\rm eph}(t^{*\prime})$ & 0 \\
         &               &               &            \\
$L_1$    & $t_o$         & $\ve{x}_{A}^{\rm eph}(t_o)$ & $\dot{\ve{x}}_{A}^{\rm eph}(t_o)$ \\
$L_2$    & $t^{\rm ca}$  & $\ve{x}_{A}^{\rm eph}(t^{\rm ca})$ & $\dot{\ve{x}}_{A}^{\rm eph}(t^{\rm ca})$ \\
\end{tabular}
\caption{
\label{Table-pN-solutions}
Choice of the constants for the six analytical post-Newtonian solutions
for the light trajectory considered in the present paper (see text for
further explanations).
}
\end{table}

\section{Simulations}
\label{Section-simulations}

\begin{figure*}
\sidecaption
\includegraphics[width=12cm]{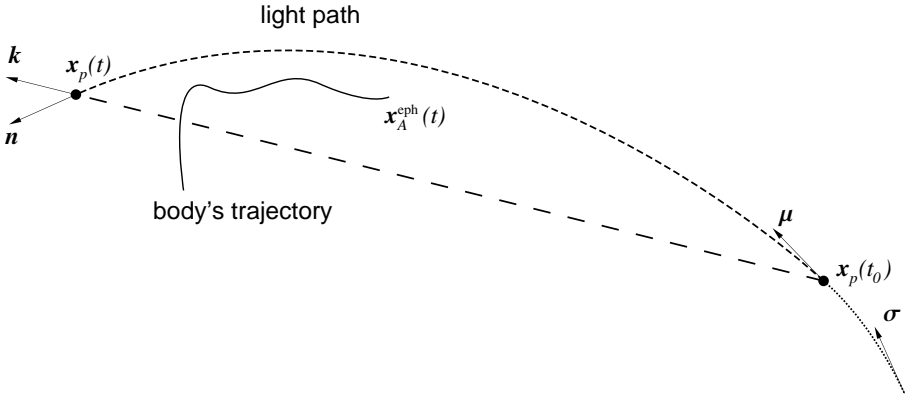}
\caption{Four vectors appearing in the calculations: vector $\ve{\mu}$
is the unit light direction at the point of emission $\ve{x}_p(t_0)$,
$\ve{n}$ is the unit light direction at the point of observation
$\ve{x}_p(t)$, $\ve{k}$ is the unit coordinate direction from
$\ve{x}_p(t_0)$ to $\ve{x}_p(t)$, and $\ve{\sigma}$ is the unit direction
of the light propagation for $t\to-\infty$. Formal definitions of these
vectors are given in Appendix \ref{Appendix-boundary-problem}.
}
\label{Figure-n-mu-k-sigma}
\end{figure*}

It is clear that numerical integration of the differential equations of
light propagation can be performed only for sources situated at some
finite distance from the gravitating body (the end point of numerical
integrations is anyway at some finite distance since it is defined by
the position of observer). Light propagation from a source at a finite
distance to the observer represents a two point boundary problem for the
differential equations of light propagation. As discussed by
\citet{Klioner:2003a} the goal of the relativistic reduction of
observations in this case is to relate the unit direction $\ve{n}$ of
the light propagation at the moment of observation to the unit
direction $\ve{k}$ from the point of light emission $\ve{x}_p(t_0)$ to
the point of light observation $\ve{x}_p(t)$ (see, Figure
\ref{Figure-n-mu-k-sigma}).

In each individual simulation we fix the points of emission and
observation for all the solutions. These points are computed by
numerical integration of the post-Minkowskian differential equations,
so that for this most accurate solution we compute vectors $\ve{n}$,
$\ve{\mu}$, $\ve{x}_p(t_0)$, $\ve{x}_p(t)$ and $\ve{k}$ with the maximal
possible accuracy. Using the formulas given in Appendix
\ref{Appendix-boundary-problem} we then solve the two point boundary
problem for all the analytical solutions discussed in our simulations
and compare the vector $\ve{n}$ from the post-Minkowskian numerical
integration to the vectors $\ve{n}$ from the analytical solutions. We
choose the distance between the point of emission and the gravitating
body sufficiently large so that the differences between the vectors
$\ve{n}$ do not change (within an accuracy of 0.001 \muas) when we
further increase the distance. Therefore, we can claim that our results
are valid also for sources at infinity. Moreover, the simulations have
shown that the differences between vectors $\ve{n}$ for the points of
emission at lower distances are smaller than those for sources at
infinity. This means that the maximal errors given below are valid also
for sources situated at finite distances from the observer (e.g. for
the solar system objects).

In order to test the analytical formulas in different situations and
check internal consistency of the simulations we have done three
independent series of simulations with different choices for the
trajectories $\ve{x}_{A}^{\rm eph}(t)$ of the bodies. For all the
simulations the masses of the gravitating bodies have been taken from
the JPL ephemeris DE405/LE405 while the radii and other parameters of
the bodies were taken from \cite{Weissman:et:al:1999}. The three series
of simulations can be described as follows.

\bigskip

{\bf A}. Parabolic trajectories with constant acceleration (Table
\ref{Table-parabolic}). For each trajectory the velocity and the
acceleration of the body at the moment of closest approach between the
body and the photon coincide with the maximal possible barycentric
velocity and acceleration of the corresponding body of the solar
system. The impact parameter for each trajectory is chosen so that at
the moment of closest approach the distance between the photon and the
body is equal to the radius of the corresponding body (expect for the
three lines in Table \ref{Table-parabolic} with fixed minimal allowed
angular distance $\psi_{\rm min}$ between the gravitating body and the
direction of light propagation as seen by the observer; the minimal
avoidance angle for each of these three bodies (Earth, Sun and Moon)
are calculated from the condition that the minimal Sun avoidance angle
is $\psi^\odot_{\rm min}=35\degr$). The distance between the observer
(satellite) and the gravitating body is taken to be the maximal
possible distance between the GAIA satellite and the corresponding body
of the solar system (some of the effects under study become larger with
increasing this distance and it is important to use the maximal
possible distance for the simulations).
For the calculation of that maximal distance
the satellite was supposed to be
situated exactly at the Lagrange point $L_2$ of the Earth-Sun system.
After fixing all these parameters the initial conditions of the photon
trajectory are still not unique and this freedom can be used to check
all possible mutual orientations of the velocity vector of the photon
and the velocity and acceleration vectors of the body at the moment of
closest approach. The directions of all these three vectors are
independent of each other. Routinely, in a coordinate system where the
direction of the velocity of the photon is fixed we check 50 uniformly
distributed directions for each of the two other vectors to find the
maximal value of the effects under investigation. In several cases we
have checked that a finer grid of mutual orientations does not lead to
any changes in the maximal differences between the models given in
Table \ref{Table-parabolic}.

\bigskip

{\bf B}. Circular coplanar trajectories (Table \ref{Table-circular}).
The observer is supposed to be situated exactly at the Lagrange point
$L_2$ of the Earth-Sun system. Each gravitating body moves along a
circular orbit with realistic semi-major axis and mean motion. All the
orbits are coplanar. All possible configurations of the observer and
the gravitating body have been checked with a step of 0.01 of the
siderial period of the corresponding body. The impact parameter of the
light ray (i.e. the minimal distance between the photon and the body)
is chosen to be not smaller than the radius of the body, but can be
larger to meet the requirement imposed by the minimal Sun avoidance
angle $\psi^\odot_{\rm min}=35\degr$. For all the bodies for which the minimal
Sun avoidance angle influences their observability (Sun, Mercury, Venus,
Earth and Moon), the simulations has been done with and without taking
the minimal Sun avoidance angle of $\psi^\odot_{\rm min}=35\degr$ into
account. For each mutual configuration of the body and the observer 36
different initial positions and velocities of the photon have been
chosen so that in the view plane of the observer the observed
directions of light are uniformly distributed around the observed
position of the gravitating body. Table \ref{Table-circular} contains
maximal differences between the models for all the light trajectories
investigated for each of the body.

\bigskip

{\bf C}. Realistic trajectories on the basis of the JPL ephemeris DE405
(Table \ref{Table-realistic}). Simulations similar to {\bf B} have been
performed using the JPL numerical ephemeris DE405/LE405 for the
trajectories of the gravitating bodies. The orbit of the observer is
taken to be a realistic Lissajous-type orbit about the $L_2$ point of
the Earth-Sun system and was computed using the algorithm suggested by
\citet{Mignard:2002}. For each body the minimal Sun avoidance angle of
35\degr\ has been taken into account while choosing the parameters of
the light rays. All mutual configurations of the observer and the
gravitating bodies between 1 January 2008 and 31 December 2020 have
been checked with a step of 1 day (the total time span covers 4749
days). The impact parameters of the light rays are taken in the same
way as for simulation {\bf B}. Also in the same way as in simulation
{\bf B} we have checked 36 different directions of the light ray
uniformly distributed relative to the line connecting the observer and
the gravitating body. We also checked in several cases that increasing
the number of the observed light directions from 36 to, say, 360, does
not change the maximal differences between the models given in Table
\ref{Table-circular}.

\bigskip

For simulations {\bf A}, {\bf B} and {\bf C} an ANSI C program has been
written. Since the effects we are looking for can be as small as
$10^{-3}$\muas$\approx 5\cdot10^{-15}$, it is not sufficient to perform
the computations using standard "double" 64-bit arithmetic which
provides only 16 decimal digits. Routinely, we have used 80-bit
arithmetic on an Intel processor (18 decimal digits). Some additional
accuracy checks have been performed using 128-bit arithmetic on an
Ultra-Sparc processor (34 decimal digits) and have shown that the
80-bit arithmetic does not produce substantial roundoff errors and is
sufficient for our purposes. Note that since the simulations took
several weeks of computing time on a Pentium III processor running at
600 MHz (about one million photon trajectories for each gravitating body
were checked), it was hardly possible to perform the simulations within
a reasonable time using a software environment emulating arithmetic
with arbitrary precision (Maple, Mathematica, etc.).

For the numerical integrations of both the post-Minkowskian and
post-Newtonian differential equations of motion we have used the
Everhart integrator described, e.g. in
\citet{Everhart:1974,Everhart:1985}. In our program the Everhard
integrators of the orders 7, 11, 13, 15, 19, 23 and 27 are implemented,
and the internal coefficients of the integrator are coded with an
accuracy consistent with the 128-bit arithmetic. This makes it possible
to perform the numerical integration with very high precision (at least,
up to 34 decimal digits) in a quite efficient way. Our investigation
showed that the number of internal iterations within the integrator can
be chosen to be as low as 1 (or at most 2) without any loss of the
resulting accuracy. This can be understood as a consequence of almost
straight trajectories of the light. The integrator of order 19 was
found to be the most efficient for our calculations. The final global
accuracy of the numerical integrations was controlled by integrating the
solution backwards. The integration is repeated automatically with
higher accuracy parameters of the integrator until the required goal
accuracy is reached.

The most time-consuming part of the calculations is the numerical
integration of the post-Minkowskian differential equations of light
propagation given in Appendix \ref{Appendix-diff-pM}). In order to
maximally speed up the numerical integration the code calculating the
right-hand side of (\ref{eqm-photon-pM:explicit}) has been optimized
using the Maple package CODEGEN \citep{Maple:1993}. This optimization
reduced the number of necessary float-point operations roughly by a
factor of 2.

\section{Results of the simulations}
\label{Section-results}

The maximal differences between the vectors $\ve{n}$ from the numerical
post-Minkowskian solution and the various analytical solutions are
summarized in Tables \ref{Table-parabolic}--\ref{Table-realistic}. The
following conclusions from the numerical simulations can be formulated:

\begin{enumerate}

\item The results of the three independent simulations are in good
agreement with each other. This serves as an internal consistency check
of the simulations.

\item The post-Minkowskian analytical model is virtually
indistinguishable from the numerical integration of the
post-Minkowskian differential equations of light propagation and leads
to errors of order 0.002 \muas.

\item As expected the naive model for the light propagation involving
the post-Newtonian analytical solution for the body being at rest at
its position at the moment of observation (the model $P_1$) is too
inaccurate and leads to errors exceeding 1 mas.
Note that our
software code for the model $P_1$ does not check if the formally
calculated impact parameter of the light rays exceeds the radius of the
body (such a check is, of course, easy to implement, but if the
software detects that the formal impact parameter is smaller than the
radius of the body, this can serve only as an evidence of the
inaccuracy of the model $P_1$ and the situation cannot be corrected
within the model $P_1$). This is the reason why the errors for the model
$P_1$ given in Tables \ref{Table-parabolic}--\ref{Table-realistic}
sometimes exceed the maximal possible gravitational light deflection
due to the corresponding body.

\item The post-Newtonian analytical solution for the body being at rest
at its position at the moment of closest approach ($P_2$) or at the
retarded moment of time ($P_3$) are virtually indistinguishable from
each other for the solar system applications (e.g., for Jupiter the
maximal difference of these models does not exceed $7.5\times10^{-4}$
\muas).

\item Any of these two models ($P_2$ and $P_3$) allows one to attain an
accuracy of $\sim 0.18$ \muas\ for the realistic trajectories of the
gravitating bodies (Table \ref{Table-realistic}). The errors of
$\sim0.75$ \muas\ appearing in the simulation with parabolic
trajectories also follow from a simplified analytical considerations
and are quoted e.g. in Table 1 of \citet{Klioner:2003a} as upper
estimates of the effects. The reason for the discrepancy
between these estimates for the parabolic trajectories and those for the
realistic trajectories were already discussed by \citet[p. 1590, above
Eq. (34)]{Klioner:2003a} where the realistic values for Jupiter and
Saturn were predicted. One can see that the predicted realistic values
and the values from Table \ref{Table-realistic} are in good agreement.

\item Simplified calculation of the retarded moment of time as given by
(\ref{t-r-s}) (model $P_3^\prime$) increases the errors in the light
deflection. The errors attains $\sim 0.3$ \muas\ for the realistic
trajectories of the gravitating bodies.
Additional simulations show that the following formula gives the
retarded moment with sufficient accuracy (this represents one Newtonian
iteration for Eq. (\ref{t-r}))

\begin{eqnarray}\label{t-r-Newton}
t^{*\prime\prime}&=&t-{|\ve{\rho}|^2\over c\,|\ve{\rho}|-\dot{\ve{x}}^{\rm eph}_A(t)\cdot\ve{\rho}},
\qquad
\ve{\rho}=\ve{x}_p(t)-\ve{x}^{\rm eph}_A(t).
\end{eqnarray}

\noindent
A solution $P_3^{\prime\prime}$ which is similar to $P_3^{\prime}$ ,
but with $t_{\rm ref}=t^{*\prime\prime}$ has the same errors (within
the level of 0.001 \muas) as the solution $P_3$ with exact value of the
retarded moment $t_{\rm ref}=t^{*}$.

\item The post-Newtonian analytical model for a body moving with a
constant velocity is indistinguishable from the post-Minkowskian model
within the accuracy of 0.002 \muas\ provided that the position and
velocity of the body on the rectilinear model trajectory of the body
coincide with the actual positions and velocities of the body at the
moment of the closest approach (model
$L_2$).

\item If the position and velocity of the body on the rectilinear
trajectory coincide with the actual positions and velocities at the
moment of observation (model $L_1$) the error exceeds 0.1 \muas.

\end{enumerate}

Let us also note that increasing the minimal Sun avoidance angle
$\psi^\odot_{\rm min}$ would reduce the errors for the Sun, Mercury,
Venus, Earth and the Moon (all the gravitating bodies which are closer
to the Sun than the observer). Therefore, the numbers in Tables
\ref{Table-parabolic}--\ref{Table-realistic} can be considered as upper
estimates for $\psi^\odot_{\rm min}\ge35\degr$.

\section{Concluding remarks}
\label{Section-conclusion}

The results of our numerical simulations are in good agreement with the
theoretical discussion by \citet{Klioner:2003a}. These results allows
one to formulate the following practical recommendations for data
processing of microarcsecond positional observations performed from the
solar system:

\begin{itemize}

\item If an accuracy of 0.2 \muas\ is sufficient one can employ the
simple post-Newtonian analytical model for the light propagation in the
gravitational field of a motionless body and substitute in that model
the actual position of the body evaluated either at the moment of
closest approach $t^{\rm ca}$ or at the retarded moment $t^*$.
If $t^*$ is used, Eq. (\ref{t-r-Newton}) can be employed to calculate
it with sufficient accuracy.

\item If an accuracy below 0.2 \muas\ is required one should use either
the full post-Minkowskian analytical model as given in Appendix
\ref{Section-pM-analytical} (neglecting the integral $\ve{g}(t_0,t)$ in
(\ref{x-pM})) or the post-Newtonian analytical model for a body moving
with a constant velocity (Appendix \ref{Section-pN-analytical-moving})
and choose the constants $\ve{x}_{A0}$ and $\ve{v}_A$ of that model to
coincide with the actual position and velocity of the body at $t^{\rm
ca}$.

\end{itemize}


\begin{table*}
\tabcolsep=5pt
\begin{tabular}{lrrrrrrrrr}
body    & $\delta$& $\delta^2$  & $pM$    & $pN, P_1$ & $pN, P_2$ & $pN, P_3$ & $pN, P^\prime_3$ & $pN, L_1$ & $pN, L_2$ \\
        &         &             &         &           &           &           &                  &           &           \\
\hline
        &         &             &         &           &           &           &                  &           &           \\
Sun$^{*}$
        &$1.76\cdot 10^6$
                  & 15.0        & 17.7    & 43.3      & 17.8      & 17.8      & 17.8             & 17.7      & 17.7      \\
Sun, $\psi^{\odot}_{\rm min}=35\degr$
        & 13600   & 0.001       & 0.001   & 0.003     & 0.002     & 0.002     & 0.002            & 0.001     & 0.001     \\
Mercury$^{*}$
        & 82.9    & 0.0         & 0.0     & 93.2      & 0.016     & 0.016     & 0.154            & 0.601     & 0.0       \\
Venus   & 493     & 0.0         & 0.0     & 812       & 0.058     & 0.058     & 0.178            & 0.357     & 0.0       \\
Earth$^{*}$
        & 574     & 0.0         & 0.0     & 8.08      & 0.058     & 0.058     & 0.058            & 0.0       & 0.0       \\
Earth, $\psi_{\rm min}=35\degr$
        & 3.64    & 0.0         & 0.0     & 0.001     & 0.001     & 0.001     & 0.001            & 0.0       & 0.0       \\
Moon$^{*}$
        & 25.9    & 0.0         & 0.0     & 1.44      & 0.003     & 0.003     & 0.003            & 0.0       & 0.0       \\
Moon, $\psi_{\rm min}=19\degr$
        & 0.086   & 0.0         & 0.0     & 0.0       & 0.0       & 0.0       & 0.0              & 0.0       & 0.0       \\
Mars    & 116     & 0.0         & 0.0     & 143       & 0.010     & 0.010     & 0.058            & 0.096     & 0.0       \\
Jupiter & 16300   & 0.001       & 0.002   & 26900     & 0.746     & 0.746     & 0.847            & 0.292     & 0.002     \\
Saturn  & 5780    & 0.0         & 0.0     & 86100     & 0.196     & 0.196     & 0.250            & 0.036     & 0.0       \\
Uranus  & 2530    & 0.0         & 0.0     & 5960      & 0.047     & 0.047     & 0.110            & 0.085     & 0.0       \\
Neptune & 2080    & 0.0         & 0.0     & 6530      & 0.050     & 0.050     & 0.106            & 0.081     & 0.0       \\
\end{tabular}
\caption{
\label{Table-parabolic}
The result of simulations using parabolic trajectories $\ve{x}^{\rm
eph}_A(t)$. All quantities are given in \muas. The second column
$\delta$ is the maximal light deflection angle for the body computed
with the post-Minkowskian differential equations of light propagation,
while $\delta^2$ is the approximate value of the post-post-Minkowskian
terms which are neglected in our considerations. The maximal errors of
the post-Minkowskian analytical solution ($pM$) and the six
post-Newtonian solutions ($pN$) are given. The meaning of $P_i$ and
$L_i$ is discussed in Section \ref{Section-ways} and summarized in
Table \ref{Table-pN-solutions}. The number "0.0" means that the actual
value of the effect is less than 0.001 \muas. The bodies designated
with a star, e.g. "Mercury$^{*}$", cannot be observed by GAIA because
of the minimal Sun avoidance angle of at least 35\degr. The minimal impact angle
$\psi_{\rm min}$ for the Sun and the Earth coincide with the minimal
Sun avoidance angle of $\psi^\odot_{\rm min}=35\degr$, while that for
the Moon is approximately calculated from simplified orbits of GAIA and
the Moon.}
\end{table*}

\begin{table*}
\tabcolsep=5pt
\begin{tabular}{lrrrrrrrrr}
body    & $\delta$& $\delta^2$  & $pM$    & $pN, P_1$ & $pN, P_2$ & $pN, P_3$ & $pN, P^\prime_3$ & $pN, L_1$ & $pN, L_2$ \\
        &         &             &         &           &           &           &                  &           &           \\
\hline
        &         &             &         &           &           &           &                  &           &           \\
Sun$^{*}$
        &$1.75\cdot 10^6$
                  & 14.8        & 17.5    & 33.3      & 17.5      &  17.5     & 17.5             & 17.5      & 17.5      \\
Sun, $\psi^{\rm Sun}_{\rm min}=35\degr$
        & 12900   & 0.001       & 0.001   & 0.002     & 0.001     &  0.001    & 0.001            & 0.001     & 0.001     \\
Mercury$^{*}$
        & 83.1    & 0.0         & 0.0     & 315       & 0.013     &  0.013    & 0.091            & 0.192     & 0.0       \\
Mercury, $\psi^{\rm Sun}_{\rm min}=35\degr$
        & 0.007   & 0.0         & 0.0     &  0.0      & 0.0       &  0.0      & 0.0              & 0.0       & 0.0       \\
Venus$^{*}$
        & 493     & 0.0         & 0.0     & 33000     & 0.058     &  0.058    & 0.156            & 0.156     & 0.0       \\
Venus, $\psi^{\rm Sun}_{\rm min}=35\degr$
        & 493     & 0.0         & 0.0     & 33300     & 0.058     &  0.058    & 0.145            & 0.144     & 0.0       \\
Earth$^{*}$
        & 574     & 0.0         & 0.0     & 13.8      & 0.0       &  0.0      & 0.0              & 0.0       & 0.0       \\
Earth, $\psi^{\rm Sun}_{\rm min}=35\degr$
        & 3.85    & 0.0         & 0.0     & 0.0       & 0.0       &  0.0      & 0.0              & 0.0       & 0.0       \\
Moon$^{*}$
        & 26.2    & 0.0         & 0.0     & 3.06      & 0.001     &  0.001    & 0.001            & 0.0       & 0.0       \\
Moon,  $\psi^{\rm Sun}_{\rm min}=35\degr$
        & 0.092   & 0.0         & 0.0     & 0.0       & 0.0       &  0.0      & 0.0              & 0.0       & 0.0       \\
Mars    & 116     & 0.0         & 0.0     & 257       & 0.006     &  0.006    & 0.033            & 0.022     & 0.0       \\
Jupiter & 16300   & 0.001       & 0.002   & 21300     & 0.139     &  0.139    & 0.202            & 0.035     & 0.002     \\
Saturn  & 5780    & 0.0         & 0.0     & 31300     & 0.020     &  0.020    & 0.035            & 0.008     & 0.0       \\
Uranus  & 2080    & 0.0         & 0.0     & 3550      & 0.003     &  0.003    & 0.009            & 0.003     & 0.0       \\
Neptune & 2530    & 0.0         & 0.0     & 3700      & 0.002     &  0.002    & 0.007            & 0.003     & 0.0       \\
\end{tabular}
\caption{
The maximal difference between the models in the simulations with
circular coplanar trajectories $\ve{x}^{\rm eph}_A(t)$ for the
gravitating bodies. All quantities are given in \muas. See caption of
Table \ref{Table-parabolic} and Section \ref{Section-simulations} for
further details.}
\label{Table-circular}
\end{table*}

\begin{table*}
\tabcolsep=5pt
\begin{tabular}{lrrrrrrrrr}
body    & $\delta$& $\delta^2$  & $pM$    & $pN, P_1$ & $pN, P_2$ & $pN, P_3$ & $pN, P^\prime_3$ & $pN, L_1$ & $pN, L_2$ \\
        &         &             &         &           &           &           &                  &           &           \\
\hline
        &         &             &         &           &           &           &                  &           &           \\
Sun     & 13000   & 0.001       & 0.001   & 0.002     & 0.001     & 0.001     & 0.001            & 0.001     & 0.001     \\
Mercury & 0.012   & 0.0         & 0.0     & 0.0       & 0.0       & 0.0       & 0.0              & 0.0       & 0.0       \\
Venus   & 493     & 0.0         & 0.0     & 30000     & 0.058     & 0.058     & 0.147            & 0.149     & 0.0       \\
Earth   & 4.75    & 0.0         & 0.0     & 0.001     & 0.0       & 0.0       & 0.0              & 0.0       & 0.0       \\
Moon    & 0.125   & 0.0         & 0.0     & 0.0       & 0.0       & 0.0       & 0.0              & 0.0       & 0.0       \\
Mars    &   116   & 0.0         & 0.0     & 385       & 0.008     & 0.008     & 0.040            & 0.025     & 0.0       \\
Jupiter & 16300   & 0.001       & 0.002   & 19600     & 0.175     & 0.175     & 0.255            & 0.038     & 0.002     \\
Saturn  &  5780   & 0.0         & 0.0     & 27900     & 0.030     & 0.030     & 0.052            & 0.008     & 0.0       \\
Uranus  &  2080   & 0.0         & 0.0     &  3550     & 0.004     & 0.004     & 0.014            & 0.003     & 0.0       \\
Neptune &  2530   & 0.0         & 0.0     &  3700     & 0.002     & 0.002     & 0.008            & 0.003     & 0.0       \\
\end{tabular}
\caption{
The maximal difference between the models in the simulations with
realistic trajectories $\ve{x}^{\rm eph}_A(t)$ of the gravitating
bodies taken from the JPL planetary ephemeris DE405/LE405. All
quantities are given in \muas. See caption of Table
\ref{Table-parabolic} and Section \ref{Section-simulations} for further
details.}
\label{Table-realistic}
\end{table*}

\appendix

\section{Equations of null geodesics}
\label{Section-null-geodesics}

Here we summarize the formulas for the null geodesics in a weak
gravitational field used in this paper. The metric tensor
$g_{\alpha\beta}$ is supposed to have the form

\begin{equation}\label{g-eta-h}
g_{\alpha\beta}=\eta_{\alpha\beta}+h_{\alpha\beta},
\end{equation}

\noindent
where

\begin{equation}\label{eta-Minkowski}
\eta_{\alpha\beta}={\rm diag}(-1,+1,+1,+1)
\end{equation}

\noindent
is the flat Minkowski metric and
$h_{\alpha\beta}=h_{\alpha\beta}(t,\ve{x})$ is the non-Galilean part of
the metric.

\subsection{Post-Newtonian approximation}

First, we use the standard {\sl post-Newtonian} assumptions on the
orders of magnitude of $h_{\alpha\beta}$ with respect to the formal
parameter $c^{-1}$:

\begin{eqnarray}\label{h-pN}
h_{00}&=&\OO2,
\nonumber\\
h_{0i}&=&\OO3,
\nonumber\\
h_{ij}&=&\OO2.
\end{eqnarray}

\noindent
Then the equations of null geodesics in the first post-Newtonian
approximation read

\begin{eqnarray}\label{eqm-h-pN}
\ddot x^i&=&{1\over 2}\,c^2\,h_{00,i}
-h_{00,k}\,\dot x^k\,\dot x^i
-\left(h_{ik,l}-{1\over 2}\,h_{kl,i}\right)\,\dot x^k\,\dot x^l
-{1\over 2}\,h_{00,t}\,\dot x^i
\nonumber\\
&&
-\left({1\over c}\,h_{0k,j}-{1\over 2c^2}\,h_{jk,t}\right)
  \,\dot x^j\,\dot x^k\,\dot x^i
-c\,\left(h_{0i,k}-h_{0k,i}\right)\,\dot x^k
\nonumber\\
&&
-h_{ik,t}\,\dot x^k+\OO2,
\phantom{\left({1\over 2c^2}\right)}
\end{eqnarray}

\noindent
where for any small latin index $i$ except for $t$ one has
$A_{,i}={\partial\over\partial x^i}\,A$ and
$A_{,t}={\partial\over\partial t} A$. Note that the velocity of the
photon $\dot x^k={\cal O}(c)$. Eq. (\ref{eqm-h-pN}) is valid in the
post-Newtonian approximation which is also called
weak-field-slow-motion approximation. The assumption $h_{0i}=\OO3$
means that the velocity of the gravitating bodies are considered to be
small with respect to $c$. Therefore, these equations cannot be used if
the velocities of the gravitating bodies are large.

\subsection{Post-Minkowskian approximation}

Second, we use the {\sl post-Minkowskian} properties of the metric
tensor and consider all the components of $h_{\alpha\beta}$ to be of
first order with respect to gravitational constant $G$:

\begin{eqnarray}\label{h-pM}
h_{00}&=&{\cal O}(G),
\nonumber\\
h_{0i}&=&{\cal O}(G),
\nonumber\\
h_{ij}&=&{\cal O}(G).
\end{eqnarray}

\noindent
No expansion in terms of $c^{-1}$ is used here. Then, one has

\begin{eqnarray}\label{eqm-h-pM}
\ddot x^i&=&{1\over 2}\,c^2\,h_{00,i}
-h_{00,k}\,\dot x^k\,\dot x^i
-\left(h_{ik,l}-{1\over 2}\,h_{kl,i}\right)\,\dot x^k\,\dot x^l
-{1\over 2}\,h_{00,t}\,\dot x^i
\nonumber\\
&&
-\left({1\over c}\,h_{0k,j}-{1\over 2c^2}\,h_{jk,t}\right)
  \,\dot x^j\,\dot x^k\,\dot x^i
-c\,\left(h_{0i,k}-h_{0k,i}\right)\,\dot x^k
\nonumber\\
&&
-h_{ik,t}\,\dot x^k-c\,h_{0i,t}
+{\cal O}({G^2}).
\phantom{\left({1\over 2c^2}\right)}
\end{eqnarray}

\noindent
The only formal difference between (\ref{eqm-h-pM}) and
(\ref{eqm-h-pN}) is the last term $-c\,h_{0i,t}$ which has first order
with respect to $G$, but is $\OO2$ and has been omitted in
(\ref{eqm-h-pN}). Eq. (\ref{eqm-h-pM}) has been derived without any
assumption on the velocity of the bodies. This equation is valid in the
first post-Minkowskian approximation which is also called weak-field
limit.

\subsection{Initial-value problem for the equations of motion}
\label{Section-initial-value-problem}

Initial value problem for the differential equations (\ref{eqm-h-pN})
and (\ref{eqm-h-pM}) can be formulated as

\begin{eqnarray}\label{BRS:eqm:photon:initials:pN}
\ve{x}(t_0)&=&\ve{x}_0,
\nonumber \\
\dot{\ve{x}}(t_0)&=&c\,\ve{\mu}\,s(t_0),
\quad \ve{\mu}\cdot\ve{\mu}=1.
\end{eqnarray}

\noindent
Function $s$ can be generally determined from the condition
of the geodesic to be a null one:

\begin{equation}\label{geodetic-null}
g_{\alpha\beta}\,k^\alpha\,k^\beta=0,\qquad
k^\alpha=\left(1,{1\over c}\,\dot{\ve{x}}\right),
\end{equation}

\noindent
which for metric (\ref{g-eta-h}) gives

\begin{equation}\label{s-general}
s={\sqrt{(1-h_{00})\,(1+h_{ij}\,\mu^i\,\mu^j)}-h_{0i}\,\mu^i\over
1+h_{ij}\,\mu^i\,\mu^j}.
\end{equation}

\noindent
Note that $s$ is a function of time and position (via the metric
components $h_{\alpha\beta}$) as well as of the direction $\ve{\mu}$,
but for a given trajectory of the photon it can be considered as
a function of time.

Taking into account the orders of magnitude of $h_{\alpha\beta}$ given
by (\ref{h-pN}) in the first post-Newtonian approximation one gets

\begin{eqnarray}\label{s:pN:pM}
s=1-{1\over 2}\,h_{00}-{1\over 2}\,h_{ij}\,\mu^i\,\mu^j
-h_{0i}\,\mu^i+\OO4.
\end{eqnarray}

It is easy to see that this expression is valid also in the first
post-Minkowskian approximation (that is, it contains all terms of
(\ref{s-general}) linear with respect to $G$).

\section{Light propagation in the post-Newtonian approximation}
\label{Section-pN}

\subsection{Metric tensor}

The non-Galilean components of the metric tensor in the post-Newtonian
approximation read

\begin{eqnarray}\label{BRS:metric}
h_{00}&=&{2\over c^2}\,w(t,\ve{x})+\OO4,
\nonumber \\
h_{0i}&=&-{4\over c^3}\,w^i(t,\ve{x})+\OO5,
\nonumber \\
h_{ij}&=&{2\over c^2}\,\delta_{ij}\,w(t,\ve{x})+\OO4.
\end{eqnarray}

\noindent
For moving mass monopoles the potentials have the form

\begin{equation}\label{w-pN}
w=\sum_A\,{GM_A\over r_A},
\end{equation}

\begin{equation}\label{w-i-pN}
w^i=\sum_A\,{GM_A\over r_A}\,\dot{x}_A^i(t),
\end{equation}

\noindent
where $r_A=|\ve{r}_A|$,

\begin{equation}\label{r-A-t}
\ve{r}_A(t,\ve{x})=\ve{x}-\ve{x}_A(t),
\end{equation}

\noindent
$\ve{x}_A(t)$ is the position of body $A$, and
$\dot{\ve{x}}_A(t)$ is the velocity of body $A$.

The metric  (\ref{BRS:metric}) with potentials (\ref{w-pN})--(\ref{w-i-pN})
coincides with the metric tensor in the Barycentric Celestial Reference System
given by the \citet{IAU:2001} (see also \citet{Soffel:et:al:2003} for a discussion).
The higher-order multipole moments of the bodies' gravitational fields
(\ref{w-pN})--(\ref{w-i-pN}) also discussed in
\citet{IAU:2001} will be ignored in this paper.

\subsection{Differential equations of light propagation}

The post-Newtonian equations of motion of a photon then read

%

\begin{eqnarray}\label{BRS:eqm:photon:motion:new}
{\ddot{\ve{x}}}&=&\sum_A {GM_A\over r_A^2}\,
\left(A_A\,\ve{n}_A+B_A\,\ve{v}+C_A\,\ve{v}_A\right)
+\OO2,
\end{eqnarray}

\begin{eqnarray}
\label{A}
A_A&=&2+\gamma-4\,\delta,
\\
\label{B}
B_A&=&4\,(1-\alpha)\,\delta-(1-\beta)\,(2+\gamma),
\\
\label{C}
C_A&=&-4\,(1-\alpha),
\end{eqnarray}

\begin{eqnarray}\label{alpha-pN}
\alpha&=&1-\ve{n}_A\cdot\ve{v},
\\
\label{beta-pN}
\beta&=&1-\ve{n}_A\cdot\ve{v}_A,
\\
\label{gamma-pN}
\gamma&=&1-\ve{v}\cdot\ve{v},
\\
\label{delta-pN}
\delta&=&1-\ve{v}\cdot\ve{v}_A,
\end{eqnarray}

\noindent
where $\ve{n}_A=\ve{r}_A/r_A$, $\ve{v}=\dot{\ve{x}}/c$,
$\ve{v}_A=\dot{\ve{x}}_A/c$.

%
%
%
%

\subsection{Initial-value problem}

From the general formulas of Section
\ref{Section-initial-value-problem} the initial-value problem for
(\ref{BRS:eqm:photon:motion:new})--(\ref{delta-pN}) reads

\begin{eqnarray}\label{BRS:eqm:photon:initials}
\ve{x}(t_0)&=&\ve{x}_0,
\nonumber \\
\dot{\ve{x}}(t_0)&=&c\,\ve{\mu}\,s(t_0),
\quad \ve{\mu}\cdot\ve{\mu}=1,
\end{eqnarray}

\noindent
and

\begin{equation}
\label{s-pN}
s(t)=1-{2\over c^2}\sum_A{GM_A\over r_A(t)}\,
\left(1-2\,\ve{\mu}\cdot\ve{v}_A(t)\right)+\OO4.
\end{equation}

\subsection{Analytical solution for a body in uniform motion}
\label{Section-pN-analytical-moving}

Two analytical solutions of Eqs.
(\ref{BRS:eqm:photon:motion:new})--(\ref{delta-pN}) and
(\ref{BRS:eqm:photon:initials})--(\ref{s-pN}) are known: 1) the
classical solution for a body at rest, i.e. $\ve{x}_A={\rm const}$; and
2) a solution for a body moving with a constant velocity. The first
solution is clearly contained in the second one. An approximate analytical
solution for bodies having a constant velocity, that is for
$\ve{x}_A(t)=\ve{x}_{A0}+\ve{v}_A\,(t-t_0)$, where $\ve{x}_{A0}={\rm
const}$ and $\ve{v}_A={\rm const}$, has been first derived by
\citet{Klioner:1989} \citep[see also][]{Klioner:Kopeikin:1992}. The
solution reads

\begin{eqnarray}\label{BRS:photon:solution:rectilinear}
\ve{x}(t)&=&\ve{x}(t_0)+c\,\ve{\mu}\,s(t_0)\,(t-t_0)
+\Delta\ve{x}(t_0,t)
\nonumber\\
&&
-\Delta\dot{\ve{x}}(t_0)\,(t-t_0),
\\ \label{BRS:dot-x-pN}
{1\over c}\,\dot{\ve{x}}(t)&=&\ve{\mu}\,s(t_0)
+{1\over c}\,\Delta\dot{\ve{x}}(t)
-{1\over c}\,\Delta\dot{\ve{x}}(t_0),
\\
\label{x-pN}
\Delta\ve{x}(t_0,t)&=&-\sum_A\,{2GM_A\over c^2}\,
\biggl(\,\ve{d}_A\,{\cal I}_A\, +\ve{g}_A\, {\cal J}_A\biggr)+\OO4,
\\
\label{vector-d-A}
\ve{d}_A&=&\vecg{\mu}\,\times\,(\ve{r}_{A0}\,\times\,\ve{g}_A),
\\
\label{vector-g-A}
\ve{g}_A&=&\vecg{\mu}-\ve{v}_A,
\\
\label{I}
{\cal I}_A&=&{1\over |\ve{g}_A|\,|\ve{r}_A|-\ve{g}_A\cdot\ve{r}_A}-
  {1\over |\ve{g}_A|\,|\ve{r}_{A0}|-\ve{g}_A\cdot\ve{r}_{A0}},
\\
\label{J}
{\cal J}_A&=&\log { |\ve{g}_A|\,|\ve{r}_A|+\ve{g}_A\cdot\ve{r}_A\over
|\ve{g}_A|\,|\ve{r}_{A0}|+\ve{g}_A\cdot\ve{r}_{A0}},
\\
\label{R-A}
\ve{r}_A(t)&=&\ve{x}(t)-\ve{x}_A(t),
\\
\label{R-A0}
\ve{r}_{A0}&=&\ve{x}(t_0)-\ve{x}_A(t_0),
\\
\label{dot-x-pN}
{1\over c}\,\Delta\dot{\ve{x}}(t)&=&-\sum_A\,{2GM_A\over c^2}\,
\biggl(\,\ve{d}_A\,{1\over c}\,
\dot{\cal I}_A\,+\ve{g}_A\,{1\over c}\,\dot{\cal J}_A\biggr)+\OO4,
\\
\label{I-dot}
{1\over c}\,\dot{\cal I}_A&=&{|\ve{g}_A|\over |\ve{r}_A|\,(|\ve{g}_A|\,|\ve{r}_A|-\ve{g}_A\cdot\ve{r}_A)},
\\
\label{J-dot}
{1\over c}\,\dot{\cal J}_A&=&{|\ve{g}_A|\over |\ve{r}_A|}.
\end{eqnarray}

This solution has two parameters $\ve{x}_{A0}$ and $\ve{v}_A$. Note
that the values of $\ve{x}_{A0}$ and $\ve{v}_A$ are arbitrary and it is
apriori unclear how to relate $\ve{x}_{A0}$ to the actual trajectory
$\ve{x}_{A}^{\rm eph}(t)$ of the body in order to minimize the errors
(see Section \ref{Section-simulations}).

The classical solution of
(\ref{BRS:eqm:photon:motion:new})--(\ref{delta-pN}) for a body at rest
can be easily restored from
(\ref{BRS:photon:solution:rectilinear})--(\ref{J-dot}) by setting
$\ve{v}_A=0$ that implies $\ve{g}_A=\ve{\mu}$ and $|\ve{g}_A|=1$.

\section{Theoretical results in the post-Minkowskian approximation}
\label{Section-pM}

\subsection{Metric tensor}

In the post-Minkowskian approximation the metric tensor of a system of
arbitrarily moving mass monopoles can be written (see
\citet{Kopeikin:Schaefer:1999,Kopeikin:Mashhoon:2002}) with retarded
potentials similar to the Lienard-Wiechert potentials which are well
known from the classical electrodynamics \citep{Jackson:1974}. The
metric reads

\begin{eqnarray}\label{BRS:metric:pM}
h_{00}&=&{2\over c^2}\, {\cal W}(t,\ve{x})+{\cal O}(G^2),
\nonumber \\
h_{0i}&=&-{4\over c^3}\, {\cal W}^{\,i}(t,\ve{x})+{\cal O}(G^2),
\nonumber \\
h_{ij}&=&{2\over c^2}\, {\cal W}^{\,ij}(t,\ve{x})+{\cal O}(G^2),
\end{eqnarray}

\begin{eqnarray}\label{cal-W}
{\cal W}(t,\ve{x})&=&\sum_A {GM_A\over r^*_A\,\beta_*}\,
\left(2\,\Gamma_*-\Gamma_*^{-1}\right)\,,
\\
\label{cal-W-i}
{\cal W}^{\,i}(t,\ve{x})&=&\sum_A
{GM_A\over r^*_A\,\beta_*}\,\Gamma_*\,
c\,{v^*_A}^i\,,
\\
\label{cal-W-ij}
{\cal W}^{\,ij}(t,\ve{x})&=&\sum_A {GM_A\over r^*_A\,\beta_*}\,
\left(\delta^{ij}\,\Gamma_*^{-1}+
2\,\Gamma_*\,{v^*_A}^i\,{v^*_A}^j
\right)\,,
\end{eqnarray}

\noindent
where $\ve{r}^*_A=\ve{x}-\ve{x}^*_A$, $r^*_A=|\ve{r}^*_A|$,
$\ve{x}^*_A=\ve{x}_A(t^*_A)$,
$\ve{v}^*_A={d\over dt^*_A}\,\ve{x}^*_A/c$, $\ve{n}^*_A=\ve{r}^*_A/r^*_A$,
$\beta_*=1-\ve{n}^*_A\cdot\ve{v}^*_A$,
$\Gamma_*={\left(1-\ve{v}^*_A\cdot\ve{v}^*_A\right)}^{-1/2}$.
Here and below in Section \ref{Section-pM} the position
$\ve{x}_A$, velocity $\dot{\ve{x}}_A$ and acceleration
$\ddot{\ve{x}}_A$ of the gravitating bodies are computed at the time moment
$t^*_A=t^*_A(t,\ve{x})$ being the retarded moment of time defined by the
following implicit equation

\begin{equation}\label{s-t}
t^*_A+{1\over c}\,r^*_A=t.
\end{equation}

Note that $r_A(t)=r^*_A\,\beta_* +{\cal O}(r_A^2/c^2)$, so that
formally ${\cal W}=w+\OO2$, ${\cal W}^{\, i}=w^i+\OO2$, ${\cal
W}^{\,ij}=\delta^{ij}\,w+\OO2$, and Eqs.
(\ref{BRS:metric})--(\ref{r-A-t}) can be easily restored from
(\ref{BRS:metric:pM})--(\ref{cal-W-ij}) within the first post-Newtonian
approximation.

\subsection{Differential equations of light propagation}
\label{Appendix-diff-pM}

Substituting (\ref{BRS:metric:pM}) into (\ref{eqm-h-pM})
one gets

\begin{eqnarray}\label{eqm-photon-pM}
{\ddot{\ve{x}}}&=&
{\cal W}_{,i}
-{2\over c^2}\,{\cal W}_{,k}\,\dot x^k\,\dot x^i
-{2\over c^2}\,{\cal W}^{\,ik}_{\ \ \ ,l}\,\dot x^k\,\dot x^l
+{1\over c^2}\,{\cal W}^{\,kl}_{\ \ \ ,i}\,\dot x^k\,\dot x^l
\nonumber\\
&&
-{1\over c^2}\,{\cal W}_{,t}\,\dot x^i
+{4\over c^4}\,{\cal W}^{\,k}_{\ \ ,j}\,\dot x^j\,\dot x^k\,\dot x^i
+{1\over c^4}\,{\cal W}^{\,jk}_{\ \ \ ,t}\,\dot x^j\,\dot x^k\,\dot x^i
\nonumber\\
&&
+{4\over c^2}\,{\cal W}^{\,i}_{\ \ ,k}\,\dot x^k
-{4\over c^2}\,{\cal W}^{\,k}_{\ \ ,i}\,\dot x^k
-{2\over c^2}\,{\cal W}^{\,ik}_{\ \ \ ,t}\,\dot x^k
\nonumber\\
&&
+{4\over c^2}\,{\cal W}^{\,i}_{\ \ ,t}
+{\cal O}(G^2).
\end{eqnarray}

Here, as usual, $A_{,t}={\partial\over\partial t}\,A$ is the partial
derivative with respect to $t$, and for any latin index $i$ with except
for $t$ one has $A_{,i}={\partial\over\partial x^i}\,A$. Since the potentials
$\cal W$, ${\cal W}^{\,i}$ and ${\cal W}^{\,ij}$ are given by Eqs.
(\ref{cal-W})--(\ref{cal-W-ij}) explicitly as functions of the retarded
time $t^*_A=t^*_A(t,\ve{x})$ one has to use

\begin{eqnarray}\label{partial-t}
{\partial\over\partial t}\,A(t,\ve{x})&=&
{\partial\,A(t^*_A,\ve{x})\over\partial t^*_A}\,{\partial t^*_A\over\partial t},
\\
\label{partial-x-i}
{\partial\over\partial x^i}\,A(t,\ve{x})&=&
{\partial\,A(t^*_A,\ve{x})\over\partial t^*_A}\,{\partial t^*_A\over\partial x^i}+
{\partial\,A(t^*_A,\ve{x})\over\partial x^i},
\end{eqnarray}

\noindent
where from (\ref{s-t})

\begin{eqnarray}\label{partial-s-t}
{\partial t^*_A\over\partial t}&=&\beta^{-1}_*,
\\
\label{partial-s-x-i}
{\partial t^*_A\over\partial x^i}&=&
-c^{-1}\,\beta^{-1}_*\,{n^*_A}^i.
\end{eqnarray}


Substituting (\ref{cal-W})--(\ref{cal-W-ij}) into (\ref{eqm-photon-pM})
and using (\ref{partial-t})--(\ref{partial-x-i}) one gets the explicit form
of the equations of motion of a photon in the first post-Minkowskian
approximation

\begin{eqnarray}\label{eqm-photon-pM:explicit}
{\ddot{\ve{x}}}&=&\sum_A {GM_A\,\Gamma_*^3\over r^{*2}_A\,\beta_*^3} \left({\cal A}_A\,\ve{n}^*_A
+{\cal B}_A\,\ve{v}
+{\cal C}_A\,\ve{v}^*_A
+{\cal D}_A\,\ve{a}^*_A
\right)
\nonumber\\
&&
+{\cal O}(G^2),
\end{eqnarray}

\begin{eqnarray}
\label{cal-A}
{\cal A}_A&=&
\left[\,\Gamma_*^{-2}\,\gamma-2\,\delta_*^2\,\right]\,
\Gamma_*^{-2}\,\left(\Gamma_*^{-2}+\varepsilon_*\right)
\nonumber\\
&&
-\left[\,\Gamma_*^{-2}\,\gamma+2\,\delta_*^2\,\right]\,
\eta_*\,\beta_*
+\,4\,\zeta_*\,\Gamma_*^{-2}\,\beta_*\,\delta_*,
\\
\label{cal-B}
{\cal B}_A&=&
\Gamma_*^{-2}\,\biggl[-\Gamma_*^{-4}\gamma
-\Gamma_*^{-2}\Bigl\{2\delta_*(2\alpha_*-\delta_*)
           +(\epsilon_*-\beta_*)\,\gamma\Bigr\}
\nonumber\\&&
\phantom{\Gamma_*^{-2}\,\biggl[}
+2\delta_*\left\{\beta_*\delta_*-\epsilon_*\,(2\alpha_*-\delta_*)\right\}
+4\zeta_*\,\beta_*(\alpha_*-\delta_*)\biggr]
\nonumber\\&&
+\eta_*\,\beta_*\left(\Gamma_*^{-2}\gamma-2\delta_*(2\alpha_*-\delta_*)\right),
\\
\label{cal-C}
{\cal C}_A&=&
\Gamma_*^{-4}\left[4\,\delta_*\,\alpha_*-\beta_*\,\gamma\right]
\nonumber\\
&&
+2\Gamma_*^{-2}\,\biggl[\,\delta_*\,\left\{\,2\,\epsilon_*\,\alpha_*-\beta_*\,\delta_*\,\right\}
-2\zeta_*\,\beta_*\,\alpha_*\,\biggr]
\nonumber\\
&&
+4\,\eta_*\,\alpha_*\,\beta_*\,\delta_*,
\\
\label{cal-D}
{\cal D}_A&=&4\,\Gamma_*^{-2}\,\alpha_*\,\beta_*\,\delta_*\,c^{-1}\,r^*_A,
\end{eqnarray}

\begin{eqnarray}\label{alpha}
\alpha_*&=&1-\ve{n}^*_A\cdot\ve{v},
\\
\label{beta}
\beta_*&=&1-\ve{n}^*_A\cdot\ve{v}^*_A,
\\
\label{gamma}
\gamma&=&1-\ve{v}\cdot\ve{v},
\\
\label{delta}
\delta_*&=&1-\ve{v}\cdot\ve{v}^*_A,
\\
\label{epsilon}
\varepsilon_*&=&(\ve{a}^*_A\cdot\ve{n}^*_A)\,c^{-1}\,r^*_A,
\\
\label{zeta}
\zeta_*&=&(\ve{a}^*_A\cdot\ve{v})\,c^{-1}\,r^*_A,
\\
\label{eta}
\eta_*&=&(\ve{a}^*_A\cdot\ve{v}^*_A)\,c^{-1}\,r^*_A,
\end{eqnarray}

\noindent
and $\ve{v}=c^{-1}\,{d\over dt}\,\ve{x}$,
$\ve{a}^*_A={d\over dt^*_A}\ve{v}^*_A=c^{-1}\,{d^2\over dt^{*2}_A}\,\ve{x}_A(t^*_A)$.

Note that if the accelerations $\ve{a}^*_A$
are considered to be of gravitational
nature (and, therefore, $\ve{a}^*_A={\cal O}(G)$) the terms in
(\ref{eqm-photon-pM:explicit}) containing $\ve{a}^*_A$ are of order
${\cal O}(G^2)$ and can be discarded in the first post-Minkowskian
approximation. However, in our theory no equations of motion of the
gravitating bodies are considered, so that the accelerations
$\ve{a}^*_A$ can be of any nature and are considered to be given. That
is why we retain the accelerations $\ve{a}^*_A$ in
(\ref{eqm-photon-pM:explicit}) in the first post-Minkowskian
approximation.

From Eqs. (\ref{eqm-photon-pM:explicit})--(\ref{eta}) one can restore
the post-Newtonian equations of motion
(\ref{BRS:eqm:photon:motion:new})--(\ref{delta-pN}) neglecting in
(\ref{eqm-photon-pM:explicit})--(\ref{eta}) all the terms formally of
order $\OO2$ (e.g., all the acceleration-dependent terms) and taking
into account that in (\ref{eqm-photon-pM:explicit})--(\ref{eta}) the
position of the bodies are calculated at the retarded moment of time
$t^*_A$ while in (\ref{BRS:eqm:photon:motion:new})--(\ref{delta-pN}) at
moment $t$. The latter circumstance gives the following relations:
$\ve{r}_A(t)=\ve{r}_A^*-r^*_A\,\ve{v}^*_A+\OO2$,
$r_A(t)=r_A^*\,\beta_*+\OO2$,
$\ve{n}(t)=(\ve{n}^*_A-\ve{v}^*_A)/\beta_*+\OO2$.

Since from (\ref{eqm-photon-pM:explicit}) it follows that

\begin{equation}\label{x-photon-Newton+G}
\ve{x}=\ve{x}_0+c\,\ve{\mu}\,(t-t_0)+{\cal O}(G),\qquad \ve{\mu}\cdot\ve{\mu}=1
\end{equation}

\noindent
in the first post-Minkowski approximation in the right-hand side of Eq.
(\ref{eqm-photon-pM:explicit}) one can put

\begin{equation}\label{x-photon-Newton}
\ve{x}=\ve{x}_0+c\,\ve{\mu}\,(t-t_0),\qquad \ve{\mu}\cdot\ve{\mu}=1
\end{equation}

\noindent
and, therefore, $\ve{v}=\ve{\mu}$, which implies, e.g., $\gamma=0$.
Formally, this makes Eq. (\ref{eqm-photon-pM:explicit})
integrable in quadratures (since the right-hand side is simply a
function of time). We prefer here not to do so and retain Eq.
(\ref{eqm-photon-pM:explicit}) in the form of a differential equation
as given by (\ref{eqm-photon-pM}).

\subsection{Initial-value problem}

Initial value problem for Eq. (\ref{eqm-photon-pM:explicit})

\begin{eqnarray}\label{photon:initials-pM}
\ve{x}(t_0)&=&\ve{x}_0,
\nonumber \\
\dot{\ve{x}}(t_0)&=&c\,\ve{\mu}\,\widetilde s(t_0),
\quad \ve{\mu}\cdot\ve{\mu}=1
\end{eqnarray}

\noindent
with

\begin{equation}\label{s-pM}
\widetilde s(t)=1-
{2\over c^2}\sum_A{GM_A\,\Gamma_*\over r^*_A\,\beta_*}\,\theta_*^2
+{\cal O}(G^2),
\end{equation}

\begin{equation}\label{theta}
\theta_*=1-\ve{\mu}\cdot\ve{v}^*_A.
\end{equation}

\subsection{Analytical solution in the post-Minkowskian approximation}
\label{Section-pM-analytical}

The explicit form of the analytical solution in the first
post-Minkowskian approximation derived by
\citet{Kopeikin:Schaefer:1999} reads

\begin{eqnarray}\label{photon:solution:pM}
\ve{x}(t)&=&\ve{x}(t_0)+c\,\ve{\mu}\,\widetilde s(t_0)\,(t-t_0)
+\widetilde\Delta\ve{x}(t_0,t)
\nonumber\\
&&
-\widetilde\Delta\dot{\ve{x}}(t_0)\,(t-t_0),
\\[5pt]
{1\over c}\,\dot{\ve{x}}(t)&=&\ve{\mu}\,\widetilde s(t_0)+
{1\over c}\,\widetilde\Delta\dot{\ve{x}}(t)-{1\over c}\,\widetilde\Delta\dot{\ve{x}}(t_0),
\\[5pt]
\label{x-pM}
\widetilde\Delta\ve{x}(t_0,t)&=&
-\sum_A\,{2GM_A\over c^2}\,\biggl(\ve{f}(t)-\ve{f}(t_0)+\ve{g}(t_0,t)\biggr)+{\cal O}(G^2),
\\ \label{F-pM}
\ve{f}(t)&=&\Gamma_*\left(\theta_*\,{\ve{\mu}\times(\ve{r}^*_A\times\ve{\mu})\over r^*_A\,\alpha_*}
-(\ve{\mu}-\ve{v}^*_A)\,\log(r^*_A\,\alpha_*)\right),
\\ \label{cal-I-pM}
\ve{g}(t_0,t)&=&\int_{t_0}^t {\Gamma_*^3\over r_A^*\,\beta_*}\,
\biggl[\,\left(\Gamma_*^2\,\zeta_*-\theta_*\,\eta_*\right)\,\ve{\mu}\times(\ve{n}^*_A\times\ve{\mu})
\nonumber\\
&&
+\alpha_*\,\log(r^*_A\,\alpha_*)\left(\eta_*\,(\ve{\mu}-\ve{v}^*_A)-\Gamma_*^{-2}\,c^{-1}\,r^*_A\,\ve{a}^*_A\right)
\biggr]\,c\,dt,
\nonumber\\
&&
\\ \label{dot-x-pM}
{1\over c}\,\widetilde\Delta\dot{\ve{x}}(t)&=&
-\sum_A\,{2GM_A\,\over c^2}\,{\Gamma_*\,\theta_*\over r^*_A\,\beta_*}\,
\biggl(\theta_*\,{\ve{\mu}\times(\ve{n}^*_A\times\ve{\mu})\over \alpha_*}
\nonumber\\&&
+(2-\theta_*)\,\ve{\mu}-2\,\ve{v}^*_A\biggr)
+{\cal O}(G^2).
\end{eqnarray}

\noindent
Here $\ve{g}(t_0,t)$ are integrals depending on the accelerations of the
bodies. The integrals can be transformed into integrals with respect to
retarded time $t^*_A$ as described in \citet{Kopeikin:Schaefer:1999}. In
the right-hand sides of (\ref{x-pM}) and (\ref{dot-x-pM}) one should
put $\ve{x}(t)=\ve{x}(t_0)+c\,\ve{\mu}\,(t-t_0)$.

Note that the solution for ${1\over
c}\,\widetilde\Delta\dot{\ve{x}}(t)$ is exact within the first
post-Minkowskian approximation (i.e. all the omitted terms are of order
$G^2$). It is remarkable that this solution can be written in such a
compact closed form for arbitrary motion of the gravitating bodies.
Note also that the solution depends only on the position and velocity
of the bodies in a single moment of time -- the retarded moment $t^*_A$
corresponding to time $t$ -- and does not depend on the previous motion
of the bodies. If all the terms which are formally of order $\OO4$
(these are the terms at least quadratic with respect to $\ve{v}_A$) are
dropped both in (\ref{dot-x-pM}) and (\ref{dot-x-pN}) the two formulas
are equivalent. One can also show that Eqs. (\ref{x-pM})--(\ref{F-pM})
agree with Eqs. (\ref{x-pN})--(\ref{J}) in the first post-Newtonian
approximation provided that $\ve{v}_A$ is considered to be constant.

Taking into account that for any function $A^i$ and
$\ve{x}(t)=\ve{x}_0+c\,\ve{\mu}\,(t-t_0)$

\begin{eqnarray}\label{A-dot-t-s}
{d\over dt}\,A^i(t,\ve{x}(t))&=&
{\partial\over\partial t}\,A^i(t,\ve{x}(t))+
{\partial\over\partial x^j}\,A^i(t,\ve{x}(t))\,c\,\mu^j
\nonumber\\
&=&{\alpha_*\over\beta_*}\,{\partial\over\partial t^*}\,A^i(t^*,\ve{x}(t))
+{\partial\over\partial x^j}\,A^i(t^*,\ve{x}(t))\,c\,\mu^j,
\end{eqnarray}

\noindent
one can prove that a time derivative of (\ref{dot-x-pM}) exactly
coincides with the post-Minkowskian equations of motion for a photon
(\ref{eqm-photon-pM:explicit})--(\ref{eta}) provided that $\ve{v}$ is
taken to be equal to $\ve{\mu}$ in the latter (this is allowed within the
first post-Minkowskian approximation as discussed above).

A different way to derive the Kopeikin-Sch\"afer solution
(\ref{photon:solution:pM})--(\ref{dot-x-pM}) for the case of
bodies moving with constant velocities is given by
\citet{Klioner:2003b} who has shown how to derive this solution
combining the post-Newtonian solution
(\ref{BRS:photon:solution:rectilinear})--(\ref{J-dot}) for a motionless
body ($\ve{v}_A=0$) with a suitable Lorentz transformation.

\section{Two point boundary value problem for the analytical solutions}
\label{Appendix-boundary-problem}

Let us consider analytical equations of light propagation
in the form valid for both analytical models discussed above

\begin{eqnarray}\label{x-generic}
\ve{x}(t)&=&\ve{x}(t_0)+c\,\ve{\mu}\,s(t_0)\,(t-t_0)
+\Delta\ve{x}(t_0,t)
\nonumber\\
&&
-\Delta\dot{\ve{x}}(t_0)\,(t-t_0)+{\cal O}(\epsilon^2),
\\ \label{dot-x-generic}
{1\over c}\,\dot{\ve{x}}(t)&=&\ve{\mu}\,s(t_0)
+{1\over c}\,\Delta\dot{\ve{x}}(t)
-{1\over c}\,\Delta\dot{\ve{x}}(t_0)+{\cal O}(\epsilon^2),
\end{eqnarray}

\noindent
with

\begin{eqnarray}
\label{condition-Delta-x}
\Delta\ve{x}(t_0,t_0)&=&0,
\\
\label{condition-Delta-dot-x}
\lim_{t\to-\infty}{1\over c}\,\Delta\dot{\ve{x}}(t)&=&0.
\end{eqnarray}

\noindent
These two conditions can be proven to be valid for both analytical
models considered above. The small parameter $\epsilon$ can be
identified with $G$ for the post-Minkowskian solution and with $c^{-2}$
for the post-Newtonian solution, so that in both solutions
$\Delta\ve{x}(t_0,t)={\cal O}(\epsilon)$ and
$c^{-1}\Delta\dot{\ve{x}}(t)={\cal O}(\epsilon)$.

Let us define several vectors

\begin{eqnarray}
\label{sigma}
\ve{\sigma}&=&\lim_{t\to-\infty}{1\over c}\,\dot{\ve{x}}(t),
\\
\label{n}
\ve{n}&=&\ve{n}(t)={\dot{\ve{x}}(t)\over |\dot{\ve{x}}(t)|},
\\
\label{k}
\ve{k}&=&\ve{k}(t_0,t)={\ve{R}(t_0,t)\over |\ve{R}(t_0,t)|},
\\
\label{ve-R}
\ve{R}&=&\ve{R}(t_0,t)=\ve{x}(t)-\ve{x}(t_0).
\end{eqnarray}

\noindent
These definitions are illustrated on Figure \ref{Figure-n-mu-k-sigma}.
Vector $\ve{\sigma}$ is the unit direction of the light path at past
null infinity ($t\to-\infty$). Vector $\ve{n}(t)$ is the unit direction
of light propagation at time moment $t$ (so that $\ve{n}(t_0)=\ve{\mu}$ and
$\lim\limits_{t\to-\infty}\ve{n}(t)=\ve{\sigma}$). Vector $\ve{k}$ is the unit
vector connecting the points $\ve{x}(t_0)$ and $\ve{x}(t)$. Now,
if one considers the points $\ve{x}(t_0)$ and $\ve{x}(t)$ as given
constants, one gets from
(\ref{x-generic})--(\ref{condition-Delta-dot-x}) the following formulas
approximately solving the two point boundary value problem for the
differential equations of motion

\begin{eqnarray}
\label{sigma-mu}
\ve{\sigma}&=&\ve{\mu}
-\ve{\mu}\times({1\over c}\,\Delta\dot{\ve{x}}(t_0)\times\ve{\mu})
+{\cal O}(\epsilon^2),
\\
\label{s-mu-dot-x}
s(t)&=&1+{1\over c}\,\ve{\mu}\cdot\Delta\dot{\ve{x}}(t)+{\cal O}(\epsilon^2),
\\
\label{n-mu}
\ve{n}&=&\ve{\mu}
+\ve{\mu}\times\left(
\left[{1\over c}\,\Delta\dot{\ve{x}}(t)-{1\over c}\,\Delta\dot{\ve{x}}(t_0)\right]
\times\ve{\mu}\right)+{\cal O}(\epsilon^2),
\\
\label{k-mu}
\ve{k}&=&\ve{\mu}
+\ve{\mu}\times\left(
\left[-{1\over c}\,\Delta\dot{\ve{x}}(t_0)
+{1\over |\ve{R}|}\,\Delta\ve{x}(t_0,t)\right]
\times\ve{\mu}\right)
\nonumber\\
&&
+{\cal O}(\epsilon^2),
\\
\label{n-k}
\ve{n}&=&\ve{k}
+\ve{k}\times\left(
\left[{1\over c}\,\Delta\dot{\ve{x}}(t)
-{1\over |\ve{R}|}\,\Delta\ve{x}(t_0,t)\right]
\times\ve{k}\right)
\nonumber\\
&&
+{\cal O}(\epsilon^2).
\end{eqnarray}

\noindent
Note that (\ref{sigma-mu}) follows from (\ref{n-mu}) for $t\to-\infty$,
that (\ref{n-k}) is a combination of (\ref{n-mu}) and (\ref{k-mu}), and
that (\ref{s-mu-dot-x}) agrees with the expressions (\ref{s-pN}) and
(\ref{s-pM}) for $s(t)$ given above for the post-Newtonian and the
post-Minkowskian solutions, respectively. From (\ref{k-mu}) and the
corresponding expressions for $\dot{\ve{x}}(t)$ and
$\Delta\ve{x}(t_0,t)$ one can see that for a given impact parameter of
the light ray the difference between vectors $\ve{k}$ and $\ve{\mu}$
becomes smaller for greater distances between the gravitating body and
point of emission $\ve{x}(t_0)$.

Note that $\Delta\ve{x}(t_0,t)$, $\Delta\dot{\ve{x}}(t)$ and
$\Delta\dot{\ve{x}}(t_0)$ on the right-hand sides of
(\ref{sigma-mu})--(\ref{n-k}) depend on $\ve{\mu}$. Formally,
considering only analytical orders of magnitude one could replace
$\ve{\mu}$ by $\ve{k}$ in these formulas. However, this works well only
when the impact parameters computing for the unperturbed trajectories
with directions $\ve{\mu}$ and $\ve{k}$ are sufficiently close to one
another. This is not always the case. For example, this assumption is
wrong in the gravitational lens limit. It is more reliable to calculate
vector $\ve{\mu}$ from given $\ve{k}$ by a numerical inversion of
(\ref{k-mu}) and then calculate $\ve{n}$ from (\ref{n-mu}).

\end{document}